\documentclass[aps,prl,twocolumn]{revtex4-1}
\usepackage{amssymb,amsmath,bm}
\usepackage{graphicx}
\usepackage{multirow}
\usepackage{hyperref}
\hypersetup{colorlinks=true,citecolor=blue,linkcolor=red,urlcolor=blue}
\usepackage{color}

\definecolor{green}{rgb}{0,0.5,0}

\newcommand{\B}[1]{{\bm{#1}}}%% Bold Roman & Greek Lower & Upper Case

\begin{document}
\title{Universal low-frequency vibrational modes in silica glasses}

\author{Silvia Bonfanti$^1$, Roberto Guerra$^1$, Chandana Mondal$^2$, Itamar Procaccia$^{2,3}$ and Stefano Zapperi$^{1,4}$ }

\affiliation{
    $^1$Center for Complexity and Biosystems, Department of Physics, University of Milan, via Celoria 16, 20133 Milano, Italy
    \\$^2$ Dept. of Chemical Physics, The Weizmann Institute of Science, Rehovot 76100, Israel
    \\$^3$Center for OPTical IMagery Analysis and Learning, Northwestern Polytechnical University, Xi'an, 710072 China
    \\$^4$ CNR - Consiglio Nazionale delle Ricerche,  Istituto di Chimica della Materia Condensata e di Tecnologie per l'Energia, Via R. Cozzi 53, 20125 Milano, Italy
}

\date{\today}

\begin{abstract}
It was recently shown that different simple models of glass formers with binary interactions define a universality class in terms of the density of states of their quasi-localized low-frequency modes. Explicitly,  once the hybridization with standard Debye (extended) modes is avoided, a number of such models exhibit a universal density of state, depending on the mode frequencies as $D(\omega) \sim \omega^4$.
It is unknown however how wide is this universality class, and whether it also pertains to more realistic models of glass formers. To address this issue we present analysis of
the quasi-localized modes in silica, a network glass which has both binary and ternary interactions. We conclude that in 3-dimensions silica exhibits the very same frequency dependence at low
frequencies, suggesting that this universal form is a generic consequence of amorphous glassiness.

\end{abstract}
\maketitle

{\bf Introduction} -- Theoretical considerations pointed out for quite some time \cite{91BGGS,03GC,03GPS,07PSG}  that low-frequency vibrational modes in amorphous glassy systems are expected to present
a density of states $D(\omega)$ with a universal dependence on the frequency $\omega$, i.e.
\begin{equation}\label{dof}
D(\omega)\sim \omega^4 \ .
\end{equation}
In spite of the fact that numerical simulations of a variety of model glass formers proliferated in recent years, the direct verification of this
prediction was late in coming. The reason for this is that the modes which are expected to exhibit this universal scaling are quasi-localized modes that
in large systems hybridize strongly with low frequency delocalized elastic (Debye) extended modes, whose density of states is expected to depend on frequency like $\omega^{d-1}$ where $d$ is the spatial dimension. To observe the universal scaling Eq.~(\ref{dof}) one needs to disentangle these types of modes.
A simple and successful idea was presented in Ref.~\onlinecite{16LDB}, using the fact that low frequency Debye modes have a lower cutoff that is determined by
the system size. By analyzing small enough systems one could isolate the relevant quasi-localized modes and their density of states, keeping the lowest
available Debye mode cleanly above the observed frequency range. Other methods were introduced to examine the density of states of the glassy modes, see e.g. Refs.~\onlinecite{shimada2018spatial,Moriel2019,angelani2018probing}.
%%%%%%%%%%%%%%%%%%%%%%%%%
\begin{figure}[b!]
	\centering
	\includegraphics[width=\columnwidth]{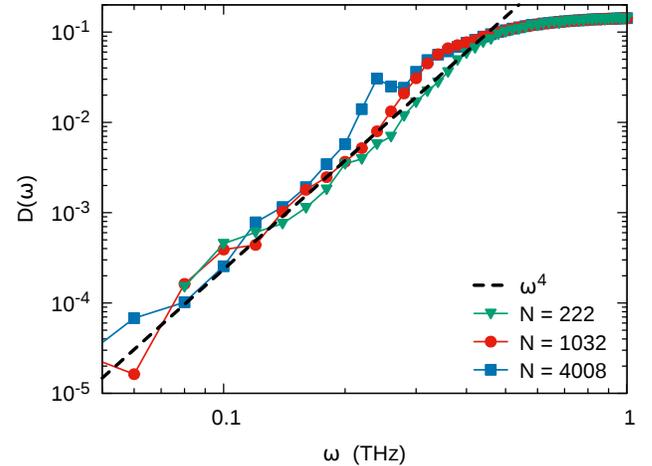}
	\caption{Density of vibrational modes $D(\omega)$ (circles) for three different system sizes. The dashed line represents the scaling law $D(\omega)\propto \omega^4$.
		One learns that the scaling law is obeyed with a diminishing range when the system size increase. It is shown below that this is due to invasion
		of the low frequency range by extended phonon modes that can hybridize with the quasi-localized modes.}
	\label{omega4}
\end{figure}
%%%%%%%%%%%%%%%%%%%%%%%%%%%%%%%%%%%%%%%%%%%%%%

Invariably, the demonstration of the universal frequency dependence Eq.~(\ref{dof}) was limited so far to models with binary interaction only. The theoretical analysis of
Refs.~\onlinecite{91BGGS,03GC,03GPS,07PSG} is, however, much more general, describing low frequency glassy modes as resulting from soft oscillators in the neighborhood of stiffer ones, and with
long-range interactions between the soft oscillators.  It is therefore timely and relevant to examine whether the universality class extends to glass formers of more realistic interactions. Here we present results for silica glass which has both binary and ternary interactions. We need to find below how to avoid the influence of low lying Debye modes, and discuss how to choose the system size to explore the density of quasi localized modes.

{\bf System and protocols} -- Our model of silica glass is simulated in 3-dimensional cubic boxes for three different system size:
\begin{itemize}
	\item $N=222$ atoms composed by $N_{Si}=74$ silicon atoms and $N_O=148$ oxygen atoms. Box length $L=15$\AA, 1000 configurations.
	\item $N=1032$ atoms composed by $N_{Si}=344$ silicon atoms and $N_O=688$ oxygen atoms. Box length $L=25$\AA , 1000 configurations.
	\item $N=4008$ atoms composed by $N_{Si}=1336$ silicon atoms and $N_O=2672$ oxygen atoms. Box length $L=39.3$\AA, 250 configurations.
\end{itemize}
The interaction between atoms is given by the Watanabe's potential~\cite{watanabe2004improved} following Refs.~\onlinecite{bonfanti2018,bonfanti2019}.
Units in the following are defined on the basis of energy, length, and time, being eV, \AA, and ps, respectively. The preparation protocol starts with randomly positioned Si,O atoms, with density ${\rho}_{in}=2.196$\,g/cm$^3$, followed by an annealing procedure:
\begin{enumerate}
	\item  After an initial 2\,ps of Newtonian dynamics with Lennard-Jones interatomic interactions, viscously damped with a rate of 1/ps and atomic velocities limited to 1\,\AA/ps, we switch to our reference Watanabe's potential for silica~\cite{watanabe2004improved}.
	\item We perform subsequent 8\,ps of damped Newtonian dynamics. iii) We then heat up the system up to 4000\,K and then quench to 0\,K in 100\,ps. Analysis on such initial samples compares well with experimentally observed density~\cite{brueckner1970properties} and with previous calculations of atomic coordination \cite{vollmayr2013temperature}. The so-produced configurations are then minimized through the fast inertial relaxation engine (FIRE) \cite{bitzek2006structural} until the total force on every atom satisfies $|\B F_i|\le 10^{-10}$\,eV/{\AA}.
\end{enumerate}

{\bf The low frequency vibrational modes} -- Denote as $U(\B r_1,\B r_2, \cdots \B r_N)$ the total potential energy of the system with $\{ \B r_i\}_{i=1}^N$ being the coordinates of the particles.
As usual  \cite{99ML,06ML,19DIP}, the modes of the system in athermal conditions ($T=0$) are obtained by diagonalizing the Hessian matrix \footnote{For systems at finite temperature special considerations are necessary, see for example \cite{19DIP}}:
\begin{equation}
H_{ij}^{\alpha\beta} \equiv \frac{\partial^2 U(\B r_1,\B r_2, \cdots \B r_N)}{\partial  r^\alpha_i \partial r^\beta_j} =-\frac{\partial F_i^\alpha}{\partial r_i^\beta} \ .
\label{Hessian}
\end{equation}
The mode frequencies~$\omega$ are obtained by the square root of the Hessian eigenvalues, and we define $\omega_{min}$ as the lowest frequency after removing the three translational zero modes.
The eigenvectors provide information on which modes are localized and which are not, as seen below.  In our simulations, the Hessian matrix is computed numerically from the first-order derivatives of inter-particle forces, cf Eq.~(\ref{Hessian}). Each element $H^{\alpha\beta}_{ij}$ is obtained by calculating the force $F_i^\alpha$ on particle $i$ resulting from a displacement of particle $j$ by a small amount, $\Delta(r_j^\beta)=10^{-7}$\,\AA~along positive and negative $\beta$-direction, and by applying the difference quotient. All the simulations have been performed using the {\small LAMMPS} simulator
package \cite{lammps}, and visualized with the {\small OVITO} package \cite{ovito}.

{\bf Results} -- In Figure~\ref{omega4} we report the density of states for the lowest frequencies in each of the three simulated system sizes. In general
we see that the predicted power law $\omega^4$ fits very well the low frequencies tail.
Interestingly, for the smallest system with $N=222$ the power law extends throughout, whereas for the larger two systems we see
the peak belonging to elastic modes sneaking in from above, invading lower frequencies for the largest system with $N=4008$. To
substantiate this, we computed the participation ratio associated with the modes in the pure power law regime and with modes whose frequency is larger than 0.3 THz.

\begin{figure}
	\centering
	\includegraphics[width=\columnwidth]{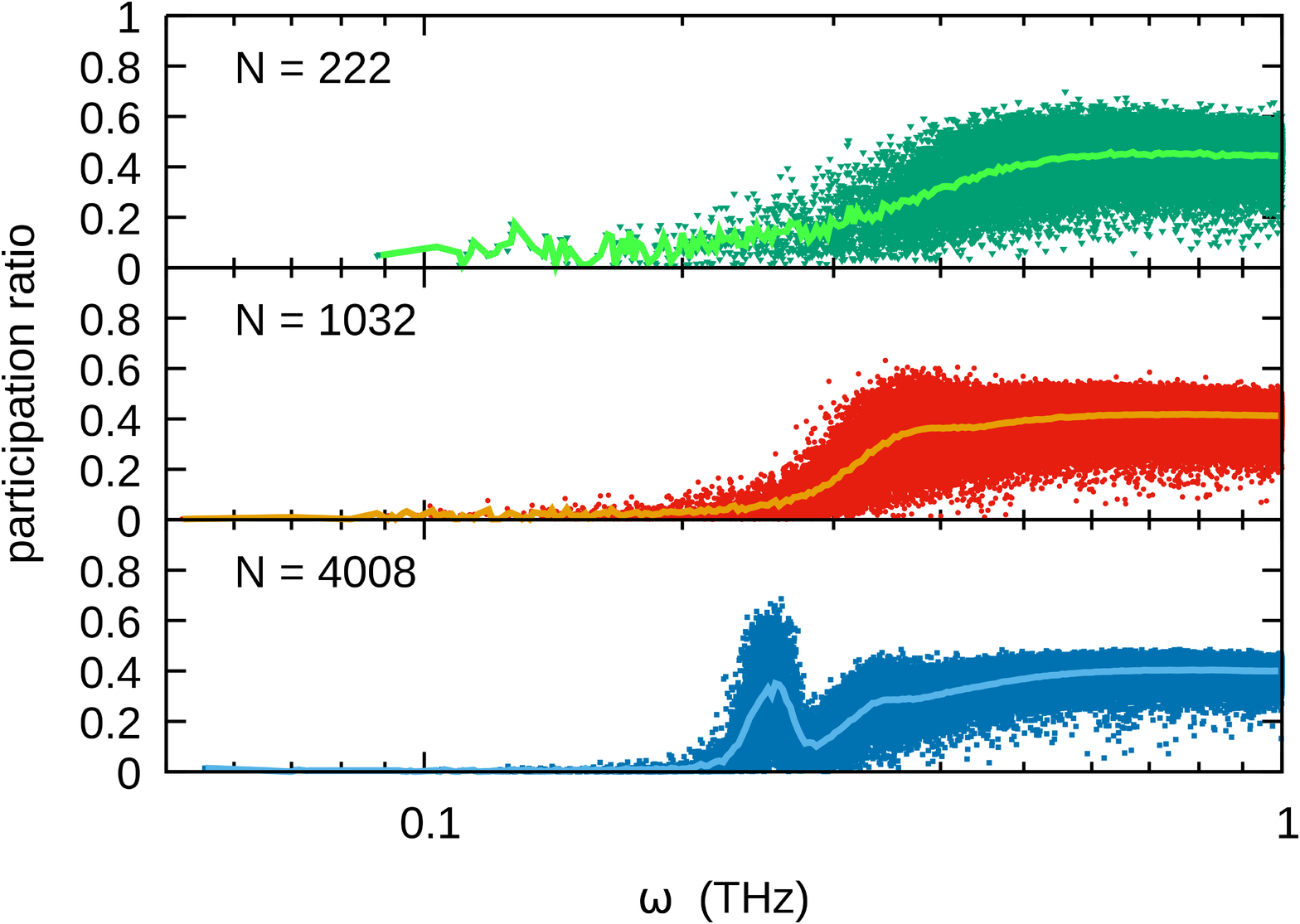}
	\caption{Participation ratio of all the modes whose frequency $\omega<1$ as a function of the frequency.
	The highlighted line is the average over the participation ratios of modes in the same band of frequencies.  }
\label{PRplot}
\end{figure}

To understand the range of frequencies for which the universal law (\ref{dof}) is expected to hold, we note that for the smallest system with
$N=222$ (cf. Fig.~\ref{dof}) this range extends up to $\omega\approx 0.4$. For the larger systems the range is smaller, up to about $\omega\approx 0.3$
for $N=1032$, becoming smallest for $N=4008$ where it ends just about $\omega\approx 0.2$. We show now that this is due to the invasion of extended modes which do not belong to the quasi-localized modes of interest. To establish this we compute the participation ratio of all the modes, and present the results in Fig.~\ref{PRplot}. The participation ratio $PR$ is defined as usual
\begin{equation}
PR= \frac{\sum_{i=1}^N(\B e_i\cdot \B e_i)^2}{[\sum_{i=1}^N(\B e_i\cdot \B e_i)]^2} \ ,
\end{equation}
where $\B e_i$ is the $i$th element of a given eigenvector of the Hessian matrix.
Localized modes are characterized by a low participation ratio, below $PR\approx 0.2$, whereas fully extended
modes have $PR=O(1)$. Examining Fig.~\ref{PRplot}, we see that for $N=222$ modes with $PR<0.2$ go all the way
to $\omega\approx 0.4$ whereas for $N=1032$ and $N=4008$ the range ends around $\omega\approx 0.3$ and $\omega\approx 0.2$
respectively. This appears to correlate very nicely with the range of scaling seen in Fig.~\ref{dof}.

\begin{figure}
	\centering
	\includegraphics[width=0.7\columnwidth]{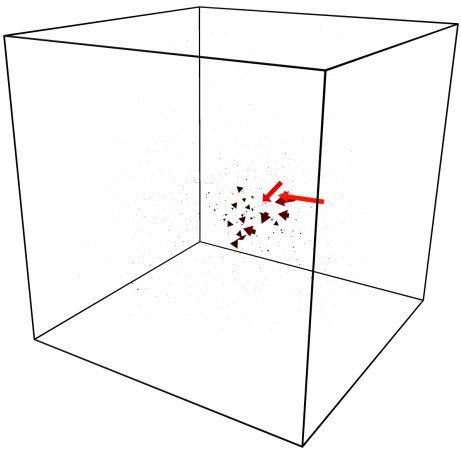}
	\caption{Orthogonal view of the eigenvector corresponding to $\omega_{min}$, for one of the $N=4008$ samples. Arrows are colored with respect to the modulus $e$ of the vectors, from black ($e=0$) to red ($e=0.6$). Arrows have been magnified by a factor of 10.
 }
\label{eigvec}
\end{figure}

An example of such localized modes, corresponding to the smallest $\omega$ value for one of the largest $N=4008$ samples, is shown in Fig.~\ref{eigvec}. This eigenvector is associated to an $\omega=0.122$\,THz, and a participation ratio $PR=0.00111 \sim 4/4008$, meaning that on average just one thousandth of the atoms is involved by this mode.

\begin{figure}
	\centering
	\includegraphics[width=\columnwidth]{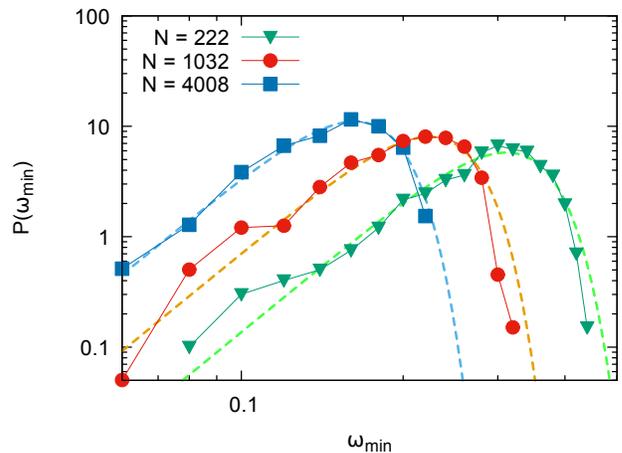}
	\caption{Distribution of the minimal vibrational frequency $P(\omega_{min})$ for the three investigated sizes. The dashed lines are the corresponding Weibull distribution  Eq.~\ref{Weib}.}
\label{scaleplot}
\end{figure}

\begin{figure}
	\centering
	\includegraphics[width=\columnwidth]{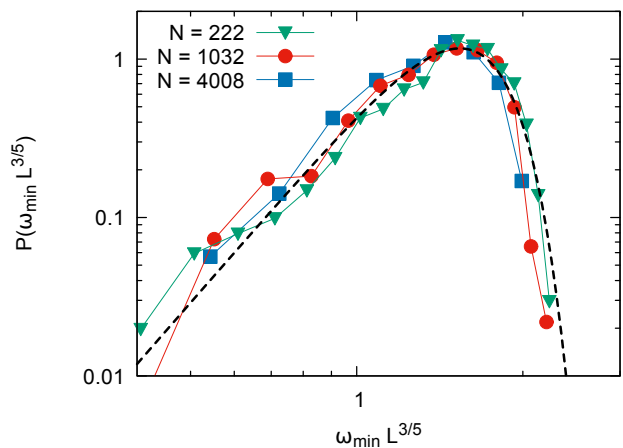}
	\caption{Distribution of the minimal vibrational frequency $P(\omega_{min})$ plotted as a function of the rescaled frequency $\omega L^\frac{3}{5}$. The continuous black line represents the Weibull distribution.}
\label{rescaled}
\end{figure}

To further solidify the universal scaling behavior of the low frequency quasi-localized modes, we turn now to extremal statistics. Since we have many configurations in our simulations, we can determine the minimal frequency obtained
from the diagonalization of the Hessian matrix in each and every configuration, denoting it as $\omega_{\rm min}$. The
average of this minimal frequency over the ensemble of configurations is $\langle \omega_{\rm min} \rangle$. Referring
to the argument first presented in Ref.~\onlinecite{10KLP}, we expect that in systems with $N$ particles,
\begin{equation}
\int_0^{\langle \omega_{\rm min}\rangle} D(\omega) d\omega \sim N^{-1} \ .
\end{equation}
Using Eq.~(\ref{dof}) we then expect that in three dimensions
\begin{equation}
\langle \omega_{\rm min} \rangle \sim N^{-1/5} \sim L^{-3/5} \ .
\label{mean}
\end{equation}
Moreover, since the different realization are uncorrelated, the values of $\omega_{\rm min}$ are also uncorrelated. Then the
celebrated Weibull theorem \cite{39Wei} predicts that the distribution of
$\omega_{\rm min}$ should obey the Weibull distribution
\begin{equation}
W(\omega_{\rm min})=\frac{5}{\langle \omega_{\rm min}\rangle^5}~\omega_{\rm min}^{4}~e^{-\left(\frac{\omega_{\rm min}}{\langle \omega_{\rm min}\rangle}\right)^5} \ .
\label{Weib}
\end{equation}
Indeed, in Fig.~\ref{scaleplot} the distribution of $\omega_{\rm min}$ for the three system size is shown, together
with the expected distribution Eq.~(\ref{Weib}). Finally, the scaling shown by Eq.~(\ref{mean}) indicates that these distribution
can be collapsed by plotting them as a function of the rescaled minimal frequency $\omega_{\rm min}L^{3/5}$.
The rescaling of the curves by $L^{3/5}$ is reported in Fig.~\ref{rescaled}.

{\bf Summary and Conclusions} -- The main aim of the Letter was to examine whether the universality class that is expressed in Eq.~(\ref{dof}) extends beyond glass formers with binary interactions. As already mentioned, quite convincing theoretical considerations predict that this universality class should be wider \cite{91BGGS,03GC,03GPS,07PSG}.
Hybridization of the glassy quasi-localized modes with regular phonon extended modes obscured for a long time the validity of Eq.~(\ref{dof}) for the former. By considering small systems
this hybridization can be avoided, exposing the universal nature of the density of states of the quasi-localized modes.
The results presented above show that a structural glass like silica, with many-body interactions much exceeding the spherical symmetry, also exhibits a dependence of the density of  quasi-localized modes on their frequency according to Eq.~(\ref{dof}) .

We note that this and other demonstrations of the universal law Eq.~(\ref{dof})  are achieved in athermal glasses at $T=0$. A separate discussion is necessary for thermal system. In that case, the configurations involved are time dependent, and there is a question on which Hessian is appropriate for describing the relevant modes.
Some ideas relevant to this question are presented in Ref.~\onlinecite{19DIP}, but the computation of the density of states remains a task for future research.

After the completion of this work, we learned of a related work by Gonzales Lopez et al
\cite{gonzalez} that supports our conclusions.

\bibliographystyle{unsrt}
\bibliography{biblio}

\begin{thebibliography}{10}

\bibitem{91BGGS}
U.~Buchenau, Yu.~M. Galperin, V.~L. Gurevich, and H.~R. Schober.
\newblock Anharmonic potentials and vibrational localization in glasses.
\newblock {\em Phys. Rev. B}, 43:5039--5045, 1991.

\bibitem{03GC}
V.~Gurarie and J.~T. Chalker.
\newblock Bosonic excitations in random media.
\newblock {\em Phys. Rev. B}, 68:134207, 2003.

\bibitem{03GPS}
V.~L. Gurevich, D.~A. Parshin, and H.~R. Schober.
\newblock Anharmonicity, vibrational instability, and the boson peak in
  glasses.
\newblock {\em Phys. Rev. B}, 67:094203, 2003.

\bibitem{07PSG}
D.~A. Parshin, H.~R. Schober, and V.~L. Gurevich.
\newblock Vibrational instability, two-level systems, and the boson peak in
  glasses.
\newblock {\em Phys. Rev. B}, 76:064206, 2007.

\bibitem{16LDB}
E.~Lerner, G.~D{\"u}ring, and E.~Bouchbinder.
\newblock {\em Phys. Rev. Lett.}, 117(3):035501, 2016.

\bibitem{shimada2018spatial}
M.~Shimada, H.~Mizuno, M.~Wyart, and A.~Ikeda.
\newblock {\em Phys. Rev. E}, 98(6):060901, 2018.

\bibitem{Moriel2019}
A.~Moriel, G.~Kapteijns, C.~Rainone, J.~Zylberg, E.~Lerner, and E.~Bouchbinder.
\newblock {\em J. Chem. Phys.}, 151(10):104503, 2019.

\bibitem{angelani2018probing}
L.~Angelani, M.~Paoluzzi, G.~Parisi, and G.~Ruocco.
\newblock {\em PNAS}, 115(35):8700--8704, 2018.

\bibitem{watanabe2004improved}
T~Watanabe, D~Yamasaki, K~Tatsumura, and I~Ohdomari.
\newblock {\em Applied surface science}, 234(1-4):207--213, 2004.

\bibitem{bonfanti2018}
Silvia Bonfanti, Ezequiel~E. Ferrero, Alessandro~L. Sellerio, Roberto Guerra,
  and Stefano Zapperi.
\newblock Damage accumulation in silica glass nanofibers.
\newblock {\em Nano Letters}, 18(7):4100--4106, 2018.
\newblock PMID: 29856226.

\bibitem{bonfanti2019}
S.~Bonfanti, R.~Guerra, C.~Mondal, I.~Procaccia, and S.~Zapperi.
\newblock Elementary plastic events in amorphous silica.
\newblock {\em Phys. Rev. E}, 100:060602, Dec 2019.

\bibitem{brueckner1970properties}
R.~Brueckner.
\newblock {\em Journal of non-crystalline solids}, 5(2):123--175, 1970.

\bibitem{vollmayr2013temperature}
K.~Vollmayr-Lee and A.~Zippelius.
\newblock {\em Physical Review E}, 88(5):052145, 2013.

\bibitem{bitzek2006structural}
E.~Bitzek, P.~Koskinen, F.~G{\"a}hler, M.~Moseler, and P.~Gumbsch.
\newblock {\em Physical review letters}, 97(17):170201, 2006.

\bibitem{99ML}
D.~L. Malandro and D.~J. Lacks.
\newblock {\em The Journal of Chemical Physics}, 110(9):4593--4601, 1999.

\bibitem{06ML}
C.~E Maloney and A.~Lema{\^\i}tre.
\newblock {\em Physical Review E}, 74(1):016118, 2006.

\bibitem{19DIP}
P.~Das, V.~Ilyin, and I.~Procaccia.
\newblock Instabilities of time-averaged configurations in thermal glasses.
\newblock {\em Phys. Rev. E}, 100:062103, 2019.

\bibitem{Note1}
For systems at finite temperature special considerations are necessary, see for
  example \cite {19DIP}.

\bibitem{lammps}
S.~Plimpton, P.~Crozier, and A.~Thompson.
\newblock Lammps-large-scale atomic/molecular massively parallel simulator.
\newblock {\em Sandia National Laboratories}, 18:43, 2007.

\bibitem{ovito}
A.~Stukowski.
\newblock {\em Modelling and Simulation in Materials Science and Engineering},
  18(1):015012, 2009.

\bibitem{10KLP}
S.~Karmakar, E.~Lerner, and I.~Procaccia.
\newblock {\em Phys. Rev. E}, 82:055103, 2010.

\bibitem{39Wei}
W.~Weibull.
\newblock {\em A Statistical Theory of the Strength of Materials}.
\newblock Generalstabens litografiska anstalts f{\"o}rlag, Stockholm, 1939.

\bibitem{gonzalez}
Karina~Gonzalez Lopez, David Richard, Geert Kapteijns, Robert Pater, and Edan
  Lerner.
\newblock Universality of the nonphononic vibrational spectrum across different
  classes of computer glasses.
\newblock {\em arXiv/3090403}, 2020.

\end{thebibliography}

\end{document}